%% file: main_v3.tex
\documentclass[%
  reprint,
nofootinbib,
nobibnotes,
 amsmath,amssymb,
aip,
graphicx,
]{revtex4-1}

\usepackage{color}
\usepackage{pdfpages}
\usepackage{pgffor}
\usepackage{xr}

\makeatletter
\AtBeginDocument{\let\LS@rot\@undefined}
\makeatother

\usepackage[section]{placeins} 
\usepackage{graphicx}
\usepackage{dcolumn}
\usepackage{bm}
\usepackage{hyperref}
\usepackage{mathtools} 
\usepackage{verbatim}
\usepackage[dvipsnames]{xcolor}
\usepackage{amsmath}
\usepackage{amsthm}
\usepackage{physics}
\usepackage[oldvoltagedirection]{circuitikz}
\usepackage{tabularx}
\usepackage[main=english, italian]{babel}

\usepackage{pgfplots}
\usepgfplotslibrary{external}
\pgfplotsset{compat=1.9}
\usepackage{tikz}
\usetikzlibrary{matrix}

\newcommand{\had}{\hat{a}^{\dagger}}
\newcommand{\ha}{\hat{a}}

\newcommand{\ista}{\affiliation{Institute of Science and Technology Austria, 3400 Klosterneuburg, Austria}}
\newcommand{\iisc}{\affiliation{Department of Instrumentation and Applied Physics$,$ Indian Institute of Science$,$ India}}
\newcommand{\iitm}{\affiliation{Department of Physics, Indian Institute of Technology Madras, Chennai 600036, India}}

\newcommand{\cquicc}{\affiliation{ Center for Quantum Information, Communication and Computing, Indian Institute of Technology Madras, Chennai 600036, India}}

\allowdisplaybreaks

\begin{document}

\title{Impedance-Engineered Josephson Parametric Amplifier with Single-Step Lithography}

\author{Lipi Patel}
\iisc
\email{lipipatel@iisc.ac.in}
\author{Samarth Hawaldar}
\iisc
\ista
\author{Aditya Panikkar}
\iisc
\author{Athreya Shankar}
\iitm
\cquicc
\author{Baladitya Suri}%
\iisc
\date{\today}

\begin{abstract}
We present the experimental demonstration of an impedance-engineered Josephson parametric amplifier (IEJPA) fabricated in a single-step lithography process. Impedance engineering is implemented using a lumped-element series LC circuit. We use a simpler lithography process where the entire device -- impedance transformer and JPA -- are patterned in a single electron beam lithography step, followed by a double-angle Dolan bridge technique for Al-AlO$_x$-Al deposition.  We observe amplification with 18 dB gain over a wide $400\,$MHz bandwidth centered around $5.3\,$GHz with added noise approaching the quantum limit, and a saturation power of $-114\,$dBm. To accurately explain our experimental results, we extend existing theories for impedance-engineered JPAs to incorporate the full sine nonlinearity of both the JPA and the transformer. Our work shows a path to simpler realization of broadband JPAs and provides a theoretical foundation for a regime of JPA operation that has been less explored in literature. 
\end{abstract}

\maketitle

Josephson Parametric Amplifiers (JPAs) \cite{yurke1989observation} have been a mainstay of superconducting circuit QED systems     \cite{vijay_quantum_jumps_2011, hartridge_backaction_2013, lin2013single, roy_amplification_2016} due to their ability to provide quantum-limited amplification\cite{caves_limits_1982}, facilitate single-shot readout\cite{krantz_single-shot_2016}, generate squeezed states of light\cite{mallet2011, yurke_squeezing_1988}, and to generate two-mode squeezed states through three-wave and four-wave mixing mechanisms\cite{eichler2011}. The phase-insensitive amplification mode of these amplifiers is usually characterized by their gain-bandwidth product defined by the maximum gain times the full width at half maximum of the amplifier. Conventional JPAs are usually narrow-band amplifiers with low gain-bandwidth products ($20\,$dB gain over $20-50\,$MHz) \cite{castellanos2007, yamamoto_fluxdriven_2008, mutus2013, zhou_nonlinear_2014}, which limit their utility to amplifying a signal composed of a single mode/tone (frequency component). On the other hand, amplifiers with higher gain-bandwidth products can be used to amplify a range of frequencies over a wider bandwidth. This finds applications in multiplexing qubit readout by allowing the amplification of multiple readout tones within the broad bandwidth of the JPA \cite{vijay2020multiplexed}.

In order to increase the bandwidth without compromising on the gain, and thereby boost the gain-bandwidth product, Travelling Wave Parametric Amplifiers (TWPAs) \cite{cullen_travelling-wave_1958, feldman_amplification_1975, wahlsten_amplification_1977, ho_eom_wideband_2012, yaakobi_amplification_2013, zorin_fluxtwpa_2019} were developed. TWPAs are transmission lines with a distributed nonlinear inductance that typically provide amplification of around $20\,$dB over a $4\,$GHz bandwidth. However, they involve complex fabrication techniques that are required to create thousands of identical unit cells with Josephson junctions (JJs) and capacitors \cite{macklin2015near, white2015traveling}. 

As simpler alternatives to the TWPA, two important techniques have been proposed and implemented to enhance the gain-bandwidth product of JPAs -- impedance-matching and impedance-engineering. Impedance-matching involves coupling the JPA oscillator, consisting of a single rf-SQUID and a shunting capacitor, to the external environment \textit{via} an impedance transformer like a Klopfenstein taper \cite{pozar} to match the impedance of the parametric oscillator (which is of the order of a few Ohms) to the $50\Omega$ environment.
However, this method also requires complex fabrication techniques to create hybrid microstrip-coplanar waveguide (CPW) transmission lines, with a large centimetre-scale footprint \cite{lu2022broadband,mutus2014strong}. The other method of increasing bandwidth, impedance engineering\cite{roy2015broadband}, involves modifying the environment seen by the JPA by introducing a transformer \textit{via} a $\lambda/2$ CPW section \cite{roy2015broadband, grebel2021flux} or a series LC resonator \cite{roy2015broadband}. Conventionally, these devices are fabricated in a multi-step lithography process, that uses a sequence of photo and e-beam lithography steps, interspersed by dielectric and metal deposition steps. 

In this Letter, we demonstrate an impedance-engineered JPA optimized for four-wave mixing-based amplification that is fabricated using a single-step lithography technique. The simplicity of our process, in turn, reduces the cost and time needed to fabricate the device. It also reduces the errors in fabrication and thereby increases the reliability and yield of the devices. The device used in this work has a simple design made up entirely of coplanar elements. Operating the device in reflection mode, we demonstrate a high gain over a wide bandwidth with noise performance approaching the quantum limit, comparable to state-of-the-art devices \cite{roy2015broadband,grebel2021flux,lu2022broadband}.  Furthermore, to explain our results, we develop a theoretical model to compute the expected gain of an impedance-engineered JPA accounting for the full non-linearity of both the JPA and the impedance transformer. Our model is a substantial extension of previous theoretical work where only the lowest order non-linearity of the JPA is considered and the transformer is assumed to be made of a linear inductor.

\begin{figure}[b]
  \includegraphics[width=0.9\linewidth]{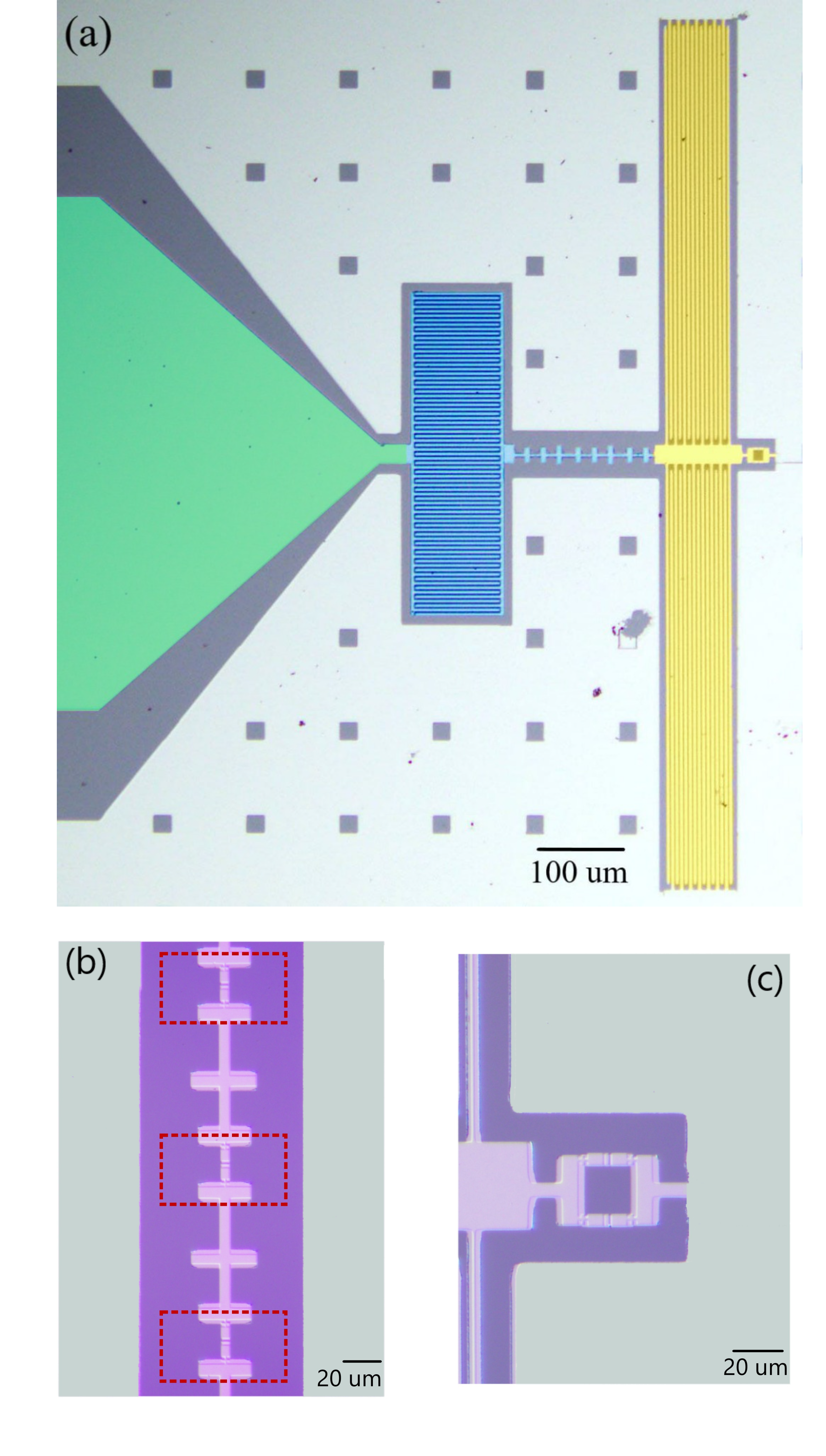}
  \caption{False-colored optical micrograph of an Impedance-engineered JPA. (a) Micrograph of the full IEJPA. From the left to right, the green colored component is the input pad, the blue colored components make up the impedance transformer, and the yellow colored elements make up the JPA. (b) Zoomed-in micrograph of the JJ array of the transformer. (c) Zoomed-in micrograph of the SQUID in the JPA.}
  \label{fig:Device}
\end{figure}

Figure~\ref{fig:Device} shows the false color micrograph of the device and the zoomed-in images of an array of JJs of the transformer and the JPA SQUID. The JPA is formed by a dc-SQUID (nominal area of $138\,\mu$m$^2$ and unbiased critical current of $1.8\,\mu$A) shunting an interdigital capacitor (IDC) with a simulated capacitance $C_J \approx 1.6\,$pF to ground. The impedance transformer, comprising an array of three nominally identical Josephson junctions, each with a critical current of $585\,$nA, emulating a nominal linear inductance of $1.69\,$nH, connected in series with another IDC (with a simulated capacitance of $C_T \approx 420\,$fF), was placed in series between the transmission line and JPA, forming a galvanic connection with both structures.  We note that a series of three JJs was used instead of a linear geometric inductor to minimize the footprint of the device. The JPA and transformer are shown in the false color micrograph in yellow and blue color respectively. 
To design the device layout, we used \verb+Qiskit Metal+ and we performed eigenmode simulations of the JPA and the impedance transformer using \verb+ANSYS HFSS+. For estimating the capacitance of the JPA and the impedance transformer, we used \verb+ANSYS HFSS+ along with \verb+COMSOL MULTIPHYSICS+. Using these simulations, in addition to the capacitances quoted above, we estimate the geometric inductance for the JPA and the transformer to be $0.06\,$nH and $0.20\,$nH, respectively.
  
The device was fabricated on an intrinsic silicon substrate using an aluminum evaporative deposition process. First, we did a preliminary cleaning of the substrate using ``piranha" (a $3:1$ volume-ratio mixture of $\text{H}_{2}\text{SO}_{4}$ and $\text{H}_{2}\text{O}_{2}$, for 10 minutes) followed by dilute hydrofluoric acid(HF) -- a $50:1$ volume-ratio mixture of DI water and HF, for 30 seconds. Next, a resist bi-layer comprising a $500\,$nm thick bottom layer of methacrylic acid (MAA) in methyl methacrylate (MMA), diluted in ethyl lactate (EL) at a 11\% concentration, and a $200\,$nm thick top layer of 4\% PMMA 950K in Anisole was spin-coated on the substrate. We then carried out e-beam lithography in a Raith e-Line at $20\,$kV acceleration voltage. We pattern both the impedance transformer and the JPA in this step. The pattern development was done using the MIBK (Methyl Isobutyl Ketone) developer. Subsequently, we used the Dolan-bridge technique for forming the Josephson junctions through the double-angle (shadow) evaporation of aluminum. The double-angle deposition was done at $\pm45^\circ$ with the first and second layers of Al having thicknesses $20\,$nm and $40\,$nm, respectively. All the Josephson junctions in the circuit -- both in the JPA SQUID loop and the impedance transformer -- were formed in a single  \textit{in-situ} oxidation step between the double-angle Al depositions. The oxidation was done at 700 mTorr pressure for 20 minutes.

The device was packaged inside a copper sample box using gold wire bonds. The sample box was then mounted on the mixing chamber plate of a dilution refrigerator at $15\,$mK. To tune the frequency of the JPA, we applied a DC magnetic flux using an external coil made of superconducting wire. The sample box containing the JPA device was mounted at the center of the external coil such that the plane of the SQUID of the JPA was perpendicular to the direction of the magnetic field of the coil. The resonance frequency $\Omega_J$ of the JPA could then be varied using the external magnetic flux $\Phi_\text{DC}$. To further prevent noise due to external magnetic fields, the sample box, along with the magnetic coil, was placed inside a superconducting aluminum shield mounted at the mixing chamber. 

\begin{figure}[tbh]
  \centering
  \includegraphics[width=\linewidth]{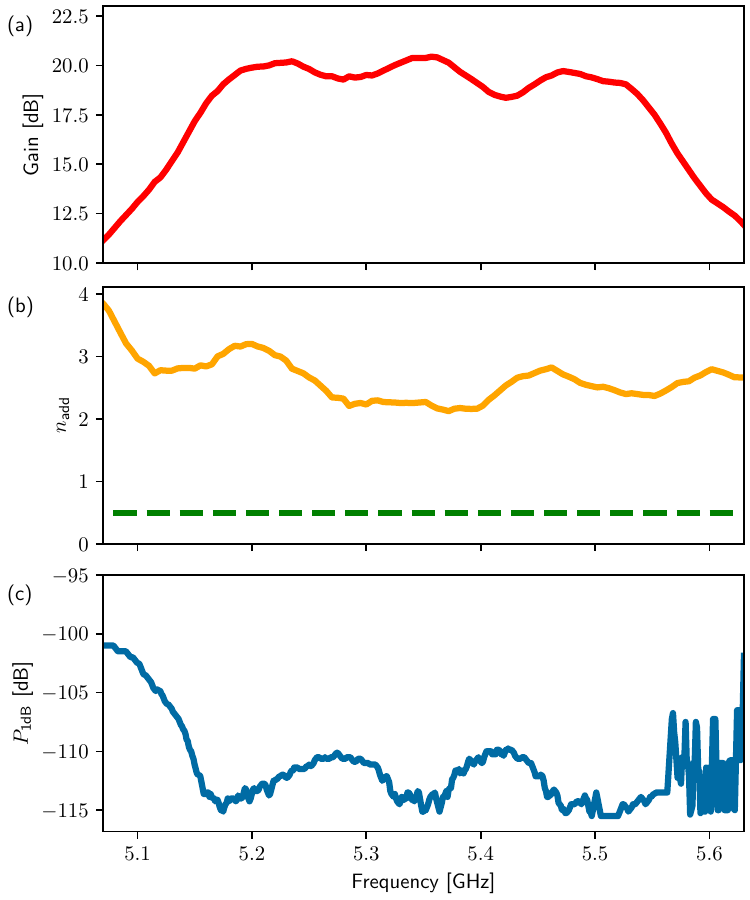}
  \caption{Measured (a) gain, (b) added noise, and (c) 1dB compression point for the IEJPA operated at a pump frequency and pump power of $5.342\,$GHz and $-88\,$dBm respectively. In the plot (b), the solid line represents the estimated $n_\text{add}$ while the dashed line represents the standard quantum limit of $0.5$ photons for a quantum limited amplifier.}
  \label{fig:jpa_measurements}
\end{figure}

We measured the gain and the added noise of the JPA by sweeping the frequency of the signal tone in steps of 1.5 MHz over a broad frequency range ($\sim\,$1 GHz) and perfoming pump on/ pump off measurements using a spectrum analyser (details of the measurement setup can be found in the Supplementary material). We optimized the gain-bandwidth product of the JPA by tuning the external DC magnetic flux, pump frequency and pump power. We obtained an optimal maximum gain of $18\,$dB over a $400\,$MHz bandwidth, at a pump frequency of $5.347\,$GHz, and a pump power of $-88\,$dBm, as shown in Fig.~\ref{fig:jpa_measurements}(a). The ripple in the gain profile can be attributed to standing waves in the cable connecting the JPA input to the slightly impedance mismatched circulator port. The ripple can be reduced by using a shorter cable between the JPA and the circulator \cite{roy2015broadband}.  

The noise photon added of the IEJPA ($n_\text{add}$) is estimated by making use of the SNR extracted from the spectrum measured using a spectrum analyser in the ``pump on" and ``pump off" cases, and using the Friis formula for noise. In summary, if we assume that the high electron-mobility transistor  (HEMT)  amplifier has a high enough gain and added noise such that the effect of noise added by  the amplifiers further in the chain can be ignored, the noise photons added, $n_\text{add}$ by the IEJPA can be estimated using
\begin{equation}
   n_\text{add} = L T_\text{HEMT} \Bigl[\frac{\text{SNR}_\text{off}}{\text{SNR}_\text{on}} - \frac{1}{G} \Bigr]\frac{k_b}{\hbar \omega},
\end{equation}
where $n_\text{add},T_\text{HEMT}$ are the added noise photons by JPA and linearised HEMT noise temperature respectively, $\text{SNR}_\text{off},\text{SNR}_\text{on}$ are the measured SNR when the pump is off and on respectively, $L$ is the loss between the JPA output and the HEMT input, and $G$ is the gain of the JPA.
In our analysis, we assume that the $T_\text{HEMT}$ is equal to the nominal noise temperature in the datasheet of $3.6\,$K. We also estimate the loss between the IEJPA and the HEMT to be about $1.8\,$dB, mostly attributed to insertion loss of the circulators and the loss from the cables.
The plot of estimated $n_\text{add}$ is shown by the orange curve in the middle panel of Fig.~\ref{fig:jpa_measurements}. We note that for a more precise characterization of the added noise, we need to use a shot-noise tunnel junction (SNTJ) noise thermometer\cite{simoen2015characterization}, which was not available during this work. The green dotted line in Fig.~\ref{fig:jpa_measurements}(b) corresponds to the standard quantum limit of $0.5\,$ photons of noise added. 

We also measured the saturation power of the JPA (also called the $1$ dB compression point), which is the power of the input signal tone at which the gain decreases by $1$ dB. To carry out this measurement, we varied the power of the input signal tone while keeping the external magnetic flux, pump power, and pump frequency at the optimal values as mentioned above. The measured saturation power was around $-114$ dBm over the bandwidth of the JPA, as shown in Fig.~\ref{fig:jpa_measurements}(c). 

As shown in Fig.~\ref{fig:jpa_model_comparision}, we find that our measured gain profile (blue, solid) is not explained sufficiently well by existing theory~\cite{roy2015broadband}, which accounts for only the lowest-order nonlinearity of the JPA  connected to a linear transformer (red dashed). Therefore, we developed a model from first-principles by deriving the full Hamiltonian of the device, writing down the quantum Langevin equations and applying input-output theory to obtain the gain expression. Importantly, our theory retains the full sine nonlinearity of the SQUID of the JPA as well as the array of JJs in the transformer. We present a detailed description of our theory in the Supplementary Material. Here, we summarize the key steps of our modeling.  

\begin{figure}[tbh]
  \centering
  \includegraphics[width=0.9\linewidth]{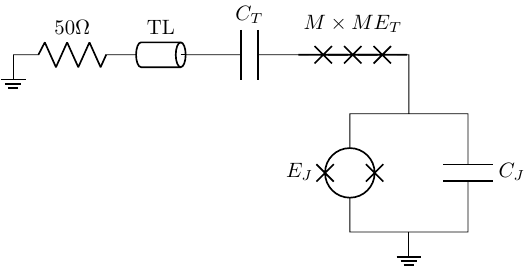}
  \caption{Equivalent circuit diagram of the device.}
  \label{fig:eq_circuit}
\end{figure}

The equivalent circuit diagram of our device is shown in Fig.~\ref{fig:eq_circuit}, where the impedance engineering is achieved by placing a series LC circuit (Transformer) between the transmission line (TL) and the JPA. In this diagram, $E_J$ and $C_J$ represent the Josephson energy and capacitance of the JPA, while $M E_T$ and $C_T$ correspond to the Josephson energy of each junction in the $M$-junction chain and capacitance of the transformer. In considering this effective circuit model we make some physically motivated simplifying assumptions which takes into account the effect of higher order nonlinearities of the device without making the device circuit too complex. We assume that for our device, the contribution from geometric inductance can be considered by just modifying the Josephson energies, or equivalently the linearized inductance, of the junctions. We justify this by noting that the geometric inductance contribution is about $15\%$ of the total linearized inductance for both the transformer and the JPA at the operating point. Another assumption that we make is that the direct capacitance from the transformer capacitance pads to ground can be neglected. This is justified from finite-element simulations that we performed that provide a capacitance to ground from the pads to be about $10\%$ of the effective capacitance between the capacitor pads, which only leads to a small frequency shift that can be accounted for by $C_T$.  The Hamiltonian for the device is 
\begin{align}
     \hat{H} &= \hat{H}_{J} + \hat{H}_{T} + \hat{H}_\text{int} + \hat{H}_\text{env}
      \nonumber\\ & = \hat{H}_\text{sys}  + \hat{H}_\text{int} + \hat{H}_\text{env}.
     \label{eqn:3}
\end{align}
Here, $\hat{H}_J$ is the JPA Hamiltonian, which includes its full sine non-linearity. It is given by 
\begin{equation}
    \hat{H}_J = -\frac{\hbar \Omega_J}{4} (\hat{A}_J - {\hat{A}_J}^{\dagger})^2 - E_J \cos( c_2 (\hat{A}_J + {\hat{A}_J}^{\dagger})),
    \label{eqn:4WM_Ham}
\end{equation}
where $\Omega_J$ is the JPA resonance frequency, $\hat{A}_J$ is the destruction operator for the intracavity mode of the JPA, $c_2 = \frac{2\pi}{\phi_0} \sqrt{\frac{\hbar Z_J}{2}}$ is the scaled zero-point phase fluctuation across the JPA junction in which $Z_J=\Omega_J L_J$ is the linearized impedance of the JPA mode with linearized inductance $L_J = \phi_0^2/4\pi^2 E_J$. The transformer Hamiltonian  $\hat{H}_T$ has the form 
\begin{align}
    \hat{H}_T &= -\frac{\hbar \Omega_{T,\text{eff}}}{4} (\hat{A}_T - {\hat{A}_T}^{\dagger})^2 \nonumber \\&  -M^2 E_T \cos( c_1 \frac{(\hat{A}_T + {\hat{A}_T}^{\dagger})}{M} -  c_2 \frac{(\hat{A}_J + {\hat{A}_J}^{\dagger})}{M}), 
    \label{eqn:4WM_Ham}
\end{align}
where the first term corresponds to the capacitor of the transformer modified by the environment and the second for an array of $M$ identical JJs. Here, $\hat{A}_T$ is the  annihilation operator for the intracavity mode of the transformer, $c_1 = \frac{2\pi}{\phi_0} \sqrt{\frac{\hbar Z_T}{2}}$ is the scaled zero-point phase fluctuation across the JJ chain in which $Z_T = \Omega_{T,\text{eff}} L_T$ is the transformer mode impedance with an effective linearized inductance $L_T=\phi_0^2/4\pi^2 E_T$, and $c_2$ is as defined above. 

The term $\hat{H}_\text{int}$ describes the interaction of the transformer with the environment and is given by 
\begin{align}
    \hat{H}_\text{int} &= \hbar \int_{0}^{\infty}\sqrt{\frac{\omega R}{2 \pi Z_T}}(\hat{A}_T(\omega) - {\hat{A}_T}^{\dagger}(\omega)) (\hat{b}(\omega) - {\hat{b}^\dagger(\omega))} \,d\omega,
    \label{eqn:7}
\end{align}
where $R$ is the impedance of the environment, and other quantities are as previously defined. In the above equation, we make an approximation by removing the explicit dependence on $\omega$ by replacing $\omega$ with $\Omega_{T,\text{eff}}$ as for the frequency ranges of interest, $\sqrt{\omega/\Omega_{T,\text{eff}}}\approx 1$. The interaction Hamiltonian can then be expressed as
\begin{equation}
    \hat{H}_\text{int} = \hbar \int_{0}^{\infty}  \sqrt{\frac{\kappa}{2 \pi}}(\hat{A}_T(\omega) - {\hat{A}_T}^{\dagger}(\omega)) (\hat{b}(\omega) - {\hat{b}^\dagger(\omega))} \,d\omega ,
    \label{eqn:8}
\end{equation}
where $ \kappa = \frac{R \Omega_{T,\text{eff}}}{Z_T}$ is the linewidth of the transformer.

The Hamiltonian $\hat{H}_\text{env}$ describing the environment is
\begin{equation}
    \hat{H}_\text{env} =\hbar \int_{0}^{\infty} \omega {\hat{b}^{\dagger}(\omega)} \hat{b}(\omega) \,d\omega ,
    \label{eqn:4WM_Ham}
\end{equation}
where $\hat{b}(\omega)$ is the annihilation operator for the TL at frequency $\omega$.

Starting from the Hamiltonian~\ref{eqn:3}, we can write down the quantum Langevin equations for the amplitudes associated with the transformer and the JPA:
\begin{equation}
    \frac{d\hat{A}_T}{dt} = -\frac{i}{\hbar}[\hat{A}_T, \hat{H}_\text{sys}] - \frac{\kappa}{2} \hat{A}_T +\sqrt{\kappa} \hat{A}_\text{in}
    \label{eqn:10}
\end{equation}
\begin{equation}
    \frac{d\hat{A}_J}{dt} = -\frac{i}{\hbar}[\hat{A}_J, \hat{H}_\text{sys}]
    \label{eqn:11}
\end{equation}

In the presence of a strong classical pump and a weak quantum signal of interest, we can break the solution into two parts by considering the classical evolution of the modes under a strong drive followed by looking at weak quantum perturbations to the mode occupation. To this end, we represent the annihilation operators of the transformer and JPA modes as $\hat{A}_T(t) = (\alpha_T + \ha_T(t))e^{-i \omega_p t}$ and  $\hat{A}_J(t) = (\alpha_J + \ha_J(t))e^{-i \omega_p t}$ where $\alpha_T, \ha_T$ and $\alpha_J, \ha_J$ are classical pump amplitudes and quantum fluctuations around them inside the transformer and JPA respectively and $\omega_p$ is the pump frequency.

First, setting the quantum fluctuations to zero ($\ha_T,\ha_J\rightarrow 0$), we obtain the equations of motion for the classical pump inside the two modes\cite{kochetov2015},
\begin{align}
    \dot{\alpha}_T &= i\omega_p \alpha_T - \left(\frac{\kappa + i\Omega_{T,\text{eff}}}{2} \right) \alpha_T \nonumber\\&+\sqrt{\kappa} \alpha_\text{in} -i \frac{E_T}{\hbar} c_1 J_1( A_\text{eff}) e^{i\phi_\text{eff}}
    \label{eqn:pump_transformer}\\
    \dot{\alpha}_J &= i\omega_p \alpha_J - i\frac{\Omega_J}{2} \alpha_J  \nonumber\\&+i \frac{E_T}{\hbar} c_2 J_1(A_\text{eff})  e^{ i\phi_\text{eff}} - i \frac{E_J}{\hbar} c_2 J_1(A_\text{jpa}) e^{ i\phi_\text{jpa}}
     \label{eqn:pump_jpa}
\end{align}
Here, $A_\text{jpa}, A_\text{eff}$ and $\phi_\text{jpa}$, $\phi_\text{eff}$ represent the amplitudes and phases of the classical pumps across the JPA and the transformer Josephson junctions respectively and are defined as
\begin{align*} 
  A_\text{eff} e^{i\phi_\text{eff}} = 2(c_1 \alpha_T -c_2 \alpha_J)/M\\ 
 A_\text{jpa} e^{i\phi_\text{jpa}} = 2(c_2 \alpha_J).
\end{align*}
These equations for the pump can be integrated starting from empty cavities to obtain the correct steady-state solutions to the pump occupation in the modes.

Next, using the steady-state solutions for the pump occupations, the equations of motion for the weak signal in the cavity can be obtained by expanding the Langevin equations to first order in the perturbations $\ha_T, \ha_J$ in the frame rotating at the pump frequecny $\omega_p$. In the frequency domain with the ``signal" frequencies being $\omega_s$, defining $\Delta = \omega_s - \omega_p$, under a rotating wave approximation, the quantum equations of motion for the intracavity perturbations are
\begin{align}
    -i\Delta\hat{a}_T(\Delta)  &=  \left(i \omega_p -i \frac{\Omega_{T,\text{eff}}}{2} - \frac{\kappa}{2} \right) \hat{a}_T(\Delta)  +\sqrt{\kappa} \hat{a}_\text{in}(\Delta) \nonumber\\ & -i\frac{E_T}{\hbar}    c_1 J_0(A_\text{eff})(c_1 \hat{a}_T(\Delta)- c_2 \hat{a}_J(\Delta)) \nonumber\\ & + i\frac{E_T}{\hbar} c_1J_2(A_\text{eff})\left(c_1 \hat{a}_T^\dagger(-\Delta)    -c_2 \hat{a}_J^\dagger(-\Delta)\right) e^{i 2\phi_\text{eff}},
    \label{eqn:weak_sig_1}\\
    -i\Delta\hat{a}_J(\Delta)  &=  i\left( \omega_p - \frac{\Omega_J}{2} \right) \ha_J(\Delta)   -i\frac{E_J}{\hbar}  c_2^2 J_0( A_\text{jpa})  \ha_J(\Delta) \nonumber\\ &   + i\frac{E_J}{\hbar} c_2^2 J_2(A_\text{jpa}) \had_J(-\Delta) e^{i 2\phi_\text{jpa}} \nonumber\\ & +i\frac{E_T}{\hbar}  c_2J_0(A_\text{eff}) (c_1 \ha_T(\Delta)- c_2 \ha_J(\Delta))  \nonumber\\ & - i\frac{E_T}{\hbar} c_2 J_2(A_\text{eff}) \left(c_1 \had_T(-\Delta)  -c_2 \had_J(-\Delta)\right)  e^{i 2\phi_\text{eff}}.  
    \label{eqn:weak_sig_2}
\end{align}
The coupled equations for \( \hat{a}_T(\Delta) \) and \( \hat{a}_J(\Delta) \) along with their Hermitian conjugates, and the input-output relation \( \sqrt{\kappa} \hat{a}_T(\Delta) = \hat{a}_\text{in}(\Delta) + \hat{a}_\text{out}(\Delta) \), can be used to derive a scattering relation of the form 
\begin{equation}
    \hat{a}_\text{out}(\Delta) = \sqrt{G(\Delta)} \hat{a}_\text{in}(\Delta) + \sqrt{G(\Delta)-1} {\hat{a}_\text{in}}^\dagger(-\Delta).
    \label{eqn:in_put eqn}
\end{equation}
In the above, $\hat{a}_\text{in}(\Delta)$ and ${\hat{a}_\text{in}}(-\Delta)$ represent the input operators for the signal and idler modes, respectively, while $\hat{a}_\text{out}(\Delta)$ is the output operator for the signal mode in the pump frame.
Here, \( G(\Delta) \) represents the gain of the signal mode and is the quantity we wish to compute from our theory model in order to compare to the experimentally measured gain profile. 

Using our theory model, we have numerically computed the gain profile as follows. First, we solve Eq.~\ref{eqn:pump_transformer} and Eq.~\ref{eqn:pump_jpa} for the steady state solution starting with $\alpha_T(t=0)=\alpha_J(t=0)=0$ with a specified input pump power, \( \alpha_\text{in} \) and a set of device parameters to obtain the long-time limit solutions $\alpha_T(t\rightarrow\infty),\alpha_J(t\rightarrow\infty)$. These steady-state pump occupations are then substituted in Eq.~\ref{eqn:weak_sig_1} and Eq.~\ref{eqn:weak_sig_2}, which we use to calculate the gain. The device parameters used in the simulations are \( \omega_p/{2 \pi} = 5.347 \text{ GHz} \), \( \Omega_J/{2 \pi} = 6.5 \text{ GHz} \), \( \Omega_{T,\text{eff}}/{2 \pi} =  6.218 \text{ GHz} \), \( L_{J} = 0.37 \text{ nH} \), and \( L_{T} = 1.9 \text{ nH} \), where the relevant inductances are the sums of linearized inductance due to the junctions and geometric inductance due to the device geometry. The extra capacitance due to coupling to the environment was calculated for the experimental setup with a $4\,$GHz bandwidth, external impedance \( R \sim 50 \, \Omega \), and \( C_T \sim 420 \, \text{fF} \). In addition, the input pump power was set to align with the experimental conditions.

Figure~\ref{fig:jpa_model_comparision} compares experimental results (blue, solid) with simulations for the IEJPA considering both finite and full nonlinearity of the Josephson junctions depicted as red dashed and green dotted lines respectively. The red theory plot considering only finite JPA nonlinearity does not match the experimental data. This observation motivated the inclusion of full junction non-linearity to better match the experimental results. The theoretical predictions considering full non-linearity predicts a reduction in the bandwidth over which appreciable gain is achievable, and hence partially explains the reduced bandwidth observed in the experiment. However, it does not match the measured profile well at frequencies far from the central frequency ($\omega_p$). This discrepancy may arise from the Markov approximation used in transitioning from Eq.\ref{eqn:7} to Eq.\ref{eqn:8}, where $\kappa$ is approximated as $\kappa = 2\pi \lambda^2({\Omega_{T,\text{eff}}})$. This approximation holds true primarily for devices with a high Q factor, which is not the case with our device. For comparison, the JPA with full sine nonlinearity without engineering is also shown in the plot as an orange dash-dotted line, where the device parameters used in the simulation are \( \omega_p/{2 \pi} = 5.347 \text{ GHz} \), \( \Omega_J/{2 \pi} = 7.05 \text{ GHz} \), and \( L_J = 0.323 \text{ nH} \). The bandwidth enhancement due to impedance engineering is clearly evident when comparing the experimentally measured gain profile to this curve. 

\begin{figure}
  \centering
  \includegraphics[width=\linewidth]{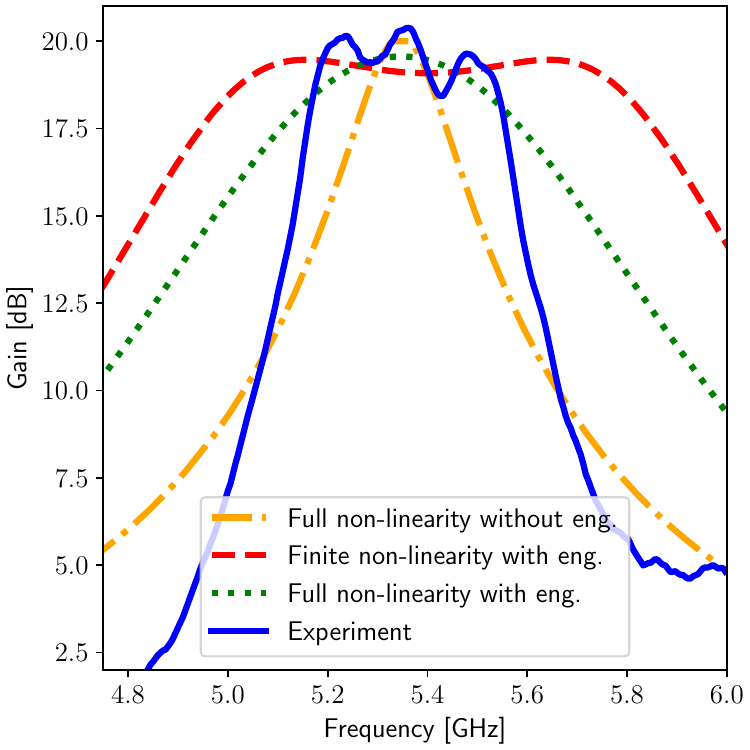}
  \caption{Comparision of experimentally measured gain with different theoretical models of the IEJPA gain for $\omega_p = 2\pi\times 5.347\,$GHz. Here, the blue solid line shows the experimentally measured gain. The non-solid lines all represent gain obtained from different models of the pumped IEJPA. The yellow dash-dotted line shows the gain obtained by considering the circuit in Fig.~\ref{fig:eq_circuit} without the transformer, but considering the full nonlinearity of the JPA junctions. The red dashed line represents the gain calculated for the IEJPA while considering only the quartic nonlinearities of the pumped junctions. Finally, the green dotted line shows the gain calculated for the pumped IEJPA while considering the full pumped junction nonlinearities up to the rotating wave approximation.}
  \label{fig:jpa_model_comparision}
\end{figure}

In summary, we have fabricated and characterized a broadband impedance-engineered JPA in a single-step fabrication procedure. We measured a gain of $18\,$dB over a 400$\,$MHz bandwidth with noise approaching the quantum limit and saturation power of -114dBm. Our experimental results are comparable to state-of-the-art values  reported in literature - impedance engineering with gain $20\,$dB, bandwidth $640\,$MHz and saturation power  $-110\,$dBm \cite{roy2015broadband} and impedance matching with gain $15\,$dB, bandwidth $700\,$MHz and saturation power  $-108\,$dBm \cite{mutus2014strong} - while using a much simpler device fabrication process. The simpler fabrication process enables the faster and more reliable realisation of near-quantum-limited broadband amplifiers for multi-qubit readout \cite{vijay2020multiplexed} or other sensing applications.
Further extensions to this work can involve improving the saturation power by using an array of SQUIDs in the JPA and increasing the number of junctions in the transformer array to decrease transformer non-linearity. Furthermore, similar devices can be designed which use three-wave (instead of four-wave) mixing, which due to the large frequency separation of the pump and signal allow the pump tone to be easily filtered without affecting the amplified signal. Such devices can hence be used as a broadband squeezed microwave source for high-precision measurements where noise in one quadrature is below vacuum noise\cite{mehmet2010}. The compact footprint of the devices makes its on-chip integration into a qubit architecture possible, thereby mitigating insertion loss due to chip-to-chip interconnects \cite{eddins2019high}.

\section*{Supplementary Material}
See the supplementary material for the details of the measurement setup for the gain and the added noise of the IEJPA. Also, we present a detailed description of our theory which includes the derivation of full Hamiltonian of the IEJPA, followed by pump and signal equations for Transformer and JPA. 
\\

The authors acknowledge the generous support of the Space Technology Cell at IISc and ISRO through the project STC-0444(2022). The authors also acknowledge the generous support of the Ministry of Electronics and Information Technology of the Govt. of India, under the centre of Excellence of Quantum Technology at the Indian Institute of Science, as well as the office of Principle Scientific Advisor, Govt. of India. SH  and AP acknowledge the support of the Kishore Vaigyanik Protsahan Yojana (KVPY). AS acknowledges the support of a New Faculty Initiation Grant (NFIG) from IIT Madras.

\section*{Data Availability}
The data that support the findings of this study are available from the corresponding author upon reasonable request.
\input{main_v3.bbl}



\section*{Supplementary material}

# Supplementary Material for "Impedance-Engineered Josephson Parametric Amplifier with Single-Step Lithography"

## 1 Measurement setup

The measurement setup for the JPA Gain and noise measurements is shown below in Fig. 1. To characterize the gain of the JPA, a strong pump tone, and a weak signal tone were applied

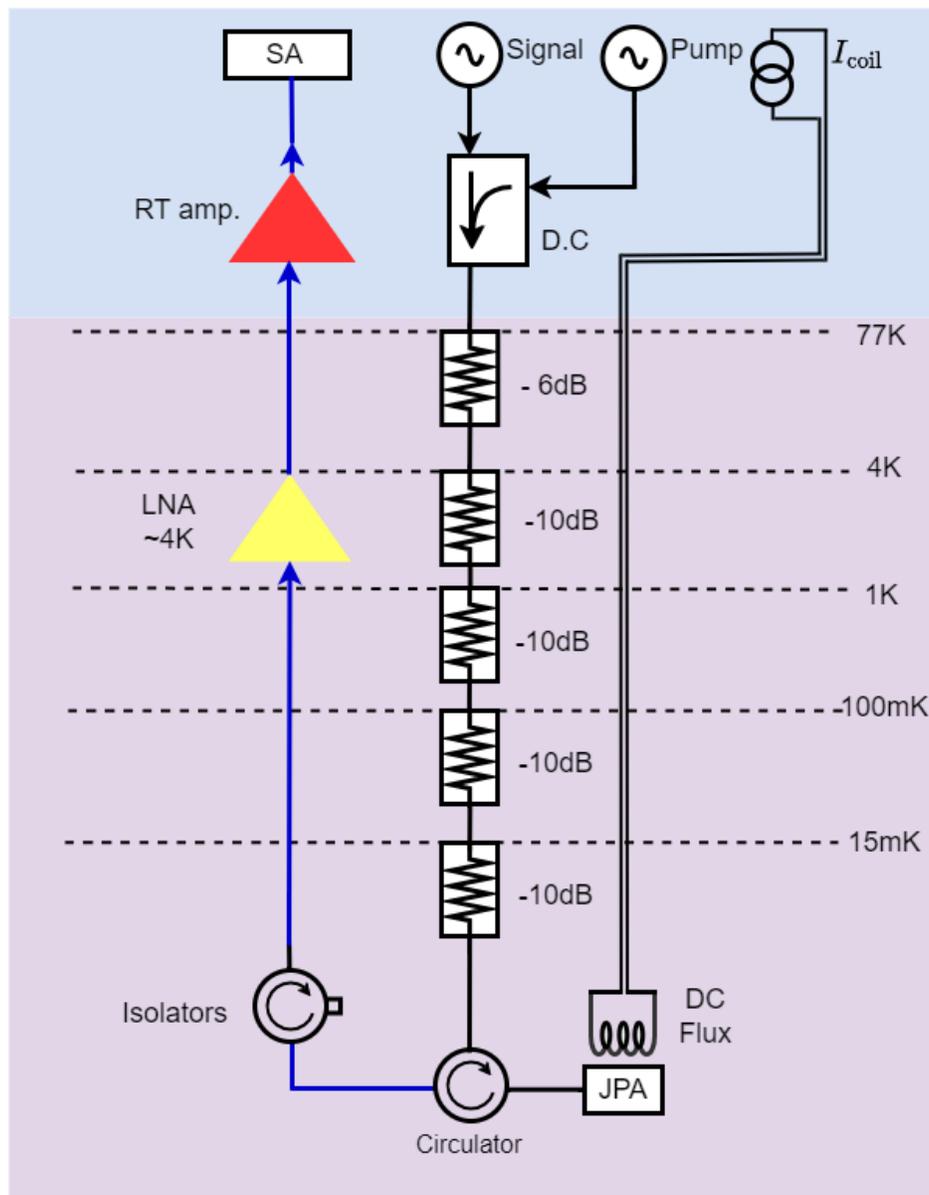

Figure 1: Schematic of the measurement setup.

along the same input coaxial line (with total attenuation of 56 dB) *via* a directional coupler (10 dB) followed by a microwave circulator (PE83CR1019) leading to the JPA. The JPA being a single-port device, the amplified signal tone that was reflected from the JPA was circulated to the low noise cryogenic HEMT amplifier (LNF-LNC0.314A). One dual-junction isolator was placed in-line between the JPA and the HEMT amplifier to effectively isolate the JPA from reflections at the HEMT amplifier. After getting amplified at the HEMT amplifier, the output signal passed through a chain of linear amplifiers at room temperature before being received by the spectrum analyser.

## 2  Full Hamiltonian of the Circuit

In this section, we derive the full Hamiltonian of the Impedance Engineered Josephson Parametric Amplifier, where engineering is implemented using a lumped-element series LC circuit.

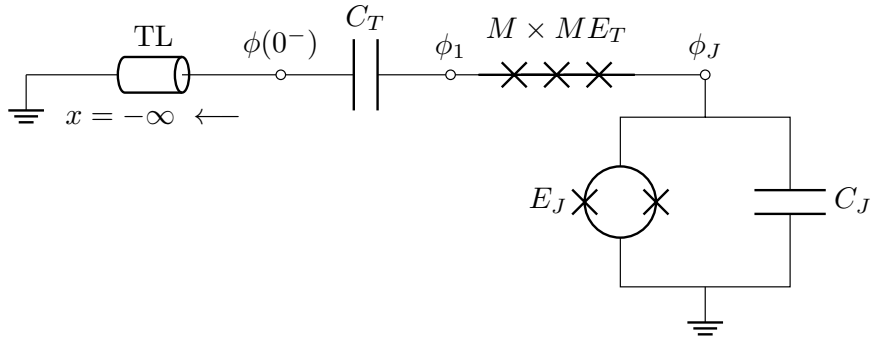

Figure 2: Equivalent circuit diagram of the device.

Fig. 2 shows the equivalent circuit diagram of the impedance-engineered JPA, where engineering is achieved by placing a series LC circuit (Transformer) between the transmission line (TL) and the JPA. In this diagram, $E_J(L_J)$ and $C_J$ represent the Josephson energy (inductance) and capacitance of the JPA with the DC-SQUID operating at some flux point, while $E_T$ and $C_T$ correspond to the Josephson energy of each of the $M$ junctions and the added capacitance of the transformer. Taking the transmission line to have inductance and capacitance per unit length of $l$ and $c$ respectively, we can write the Lagrangian of the circuit

as,

$$\mathcal{L} = \mathcal{L}_{\text{TL}} + \mathcal{L}_t + \mathcal{L}_J, \tag{1}$$

$$\mathcal{L}_{TL} = \frac{1}{2} \int_{-\infty}^{0^-} \mathrm{d}x \left( c\dot{\phi}^2(x,t) - \frac{1}{l}\phi'^2(x,t) \right), \tag{2}$$

$$\mathcal{L}_J = \frac{1}{2} C_J \dot{\phi}_J^2 + E_J \cos\left( \frac{2\pi \phi_J}{\phi_0} \right), \tag{3}$$

$$\mathcal{L}_t = \frac{1}{2} C_t (\dot{\phi}_1 - \dot{\phi}(0,t))^2 + M E_T \cos\left( \frac{\phi_J - \phi_1}{M} \right), \tag{4}$$

Note that in our system, we have used $M$ as the number of identical JJs in the array of the transformer, while scaling the Josephson energy of each junction in the array to $ME_T$. This results in effectively having a Josephson element with a reduced nonlinearity as the phase across the array is distributed equally over each junction $((\phi_1 - \phi_J) \to \frac{(\phi_1 - \phi_J)}{M})$ while maintaining the same linearised inductance[1].

This Lagrangian can be transformed to the $k$-space by substituting [2, 3]

$$\phi(x) = \int_0^\infty \mathrm{d}k\, C_1 (q(k) \cos(kx) + s(k) \sin(kx)), \tag{5}$$

where $k$ is the wavevector of stationary waves along the transmission line, and $q(k)$ and $r(k)$ are the cosine and sine quadratures of the corresponding stationary waves supported by the semi-infinite transmission line.

Under this substitution, we can further express

$$\int_{-\infty}^{0^-} \mathrm{d}x\, \dot{\phi}^2(x,t) = \int_{-\infty}^{0^-} \mathrm{d}x \left( \int_0^\infty \mathrm{d}k\, C_1(q(k)\cos(kx) + s(k)\sin(kx)) \right)^2$$

$$= C_1^2 \int_0^\infty \mathrm{d}k_1 \int_0^\infty \mathrm{d}k_2\, \frac{\pi}{2} \Big( \dot{q}(k_1)\dot{q}(k_2)\big(\delta(k_1+k_2) + \delta(k_1-k_2)\big)$$

$$+ \dot{s}(k_1)\dot{s}(k_2)\big(\delta(k_1-k_2) - \delta(k_1+k_2)\big) \Big)$$

$$= \frac{\pi}{2} \int_0^\infty \mathrm{d}k\, C_1^2 \big(\dot{q}(k)^2 + \dot{s}(k)^2\big), \tag{6}$$

and similarly

$$\int_{-\infty}^{0^-} \mathrm{d}x \frac{1}{l} \phi'^2(x,t) = \frac{\pi}{2} \frac{k^2}{l} \int_0^\infty \mathrm{d}k\, C_1^2 \big(q(k)^2 + s(k)^2\big). \tag{7}$$

3

The above expressions allow us to rewrite the Lagrangian in the $k$-space as

$$\mathcal{L}_{TL} = \frac{1}{2}\int_0^\infty dk\, \frac{\pi}{2}C_1^2\left(c\big(\dot{q}(k)^2+\dot{s}(k)^2\big) - \frac{k^2}{l}\big(q(k)^2+s(k)^2\big)\right), \tag{8}$$

$$\mathcal{L}_J = \frac{1}{2}C_J\dot{\phi}_J^2 + E_J\cos\left(\frac{2\pi\phi_J}{\phi_0}\right), \tag{9}$$

$$\mathcal{L}_t = \frac{1}{2}C_t\left(\dot{\phi}_1 - C_1\int_0^\infty dk\, \dot{q}(k)\right)^2 + ME_T\cos\left(\frac{\phi_J-\phi_1}{M}\right). \tag{10}$$

At this stage, we notice that the "sine"-part of the waves in the TL do not couple to the transformer-JPA system and evolves independently. This allows us to drop the $s(k)$ degrees of freedom from our analysis.

Further using the dispersion relation of $\omega = vk$ for $v = 1/\sqrt{lc}$, and rescaling the generalised positions $q(\omega) \to q(\omega)\sqrt{\frac{2v}{\pi C_1^2|\omega|c}}$ to absorb the dependence of density of states in the waveguide with frequency, the Lagrangian can be rewritten to

$$\mathcal{L}_{TL} = \frac{1}{2}\int_0^\infty d\omega\left(\frac{\dot{q}(\omega)^2}{\omega} - \omega q(\omega)^2\right), \tag{11}$$

$$\mathcal{L}_J = \frac{1}{2}C_J\dot{\phi}_J^2 + E_J\cos\left(\frac{2\pi\phi_J}{\phi_0}\right), \tag{12}$$

$$\mathcal{L}_t = \frac{1}{2}C_t\left(\dot{\phi}_1 - \sqrt{2R_p}\int_0^\infty d\omega\, \frac{\dot{q}(\omega)}{\sqrt{|\omega|}}\right)^2 + ME_T\cos\left(\frac{\phi_J-\phi_1}{M}\right), \tag{13}$$

where $R = \sqrt{l/c}$ is the impedance of the transmission line, and $R_p = \frac{R}{\pi}$ is a rescaled impedance.

As we want to come to a Hamiltonian for the system, we write the moments as

$$\frac{\partial \mathcal{L}}{\partial \dot{\phi}_J} = Q_J = C_J\dot{\phi}_J, \tag{14}$$

$$\frac{\partial \mathcal{L}}{\partial \dot{\phi}_1} = Q_1 = C_t\left(\dot{\phi}_1 - \sqrt{2R_p}\int_0^\infty d\omega\, \frac{\dot{q}(\omega)}{\sqrt{|\omega|}}\right), \tag{15}$$

$$\frac{\partial \mathcal{L}}{\partial \dot{q}(\omega)} = p(\omega) = \frac{\dot{q}(\omega)}{\omega} - C_t\left(\dot{\phi}_1 - \sqrt{2R_p}\int_0^\infty d\omega\, \frac{\dot{q}(\omega)}{\sqrt{|\omega|}}\right)\cdot\sqrt{\frac{2R_p}{\omega}}. \tag{16}$$

As the limit for the frequency integral is from 0 to $\infty$, the further analysis without corrections leads to the "ultraviolet catastrophe". To avoid that, we introduce the cutoff function $h(\omega)$ such that $\int_0^\infty d\omega\, h(\omega) = K < \infty$, where $K$ can be thought of as the bandwidth

of the setup [1]. With the cutoff in place, the affected equations become,

$$Q_1 = C_t \left( \dot{\phi}_1 - \sqrt{2R_p} \int_0^\infty d\omega\, h(\omega) \frac{\dot{q}(\omega)}{\sqrt{|\omega|}} \right), \tag{17}$$

$$p(\omega) = \frac{\dot{q}(\omega)}{\omega} - C_t \left( \dot{\phi}_1 - \sqrt{2R_p} \int_0^\infty d\omega\, h(\omega) \frac{\dot{q}(\omega)}{\sqrt{|\omega|}} \right) \sqrt{\frac{2R_p}{\omega}}. \tag{18}$$

Backsubstituting $\dot{q}$ from Eqn. 17 into Eqn. 18 and solving, we can express the integrals in the momentum variables as

$$I_q = \int_0^\infty d\omega\, h(\omega) \frac{\dot{q}(\omega)}{\sqrt{|\omega|}} = \frac{\int_0^\infty d\omega\, h(\omega)\sqrt{\omega}p(\omega) + C_t\sqrt{2R_p}K\dot{\phi}_1}{1 + 2KR_pC_t} = \frac{I_p + C_t\sqrt{2R_p}K\dot{\phi}_1}{1 + 2KR_pC_t}. \tag{19}$$

Note here that had we not used the cutoff function, or equivalently used a constant cutoff such that $K \to \infty$, the infinities that have turned up would be notoriously hard to resolve.

The expression in Eqn. 19 can be used to now get a velocity-momentum relation for $\phi_1$ and $q(\omega)$ by substituting the values into Eqns. 17 and 18. As the penultimate step to having the full hamiltonian, we get the relations as

$$\dot{\phi}_1 = \frac{Q_1}{C_t} + \sqrt{2R_p}I_p + 2KR_pQ_1, \tag{20}$$

$$\dot{q}(\omega) = \omega p(\omega) + \sqrt{2R_p\omega}Q_1. \tag{21}$$

Finally, we now have the Hamiltonian of the system

$$\begin{aligned}
\mathcal{H} &= \int_0^\infty d\omega\, p(\omega)\dot{q}(\omega) + Q_1\dot{\phi}_1 + Q_J\dot{\phi}_J - \mathcal{L} \\
&= \int_0^\infty d\omega\, \omega(p(\omega)^2 + q(\omega)^2) + \sqrt{2R_p}Q_1 \int_0^\infty d\omega\, \sqrt{\omega}p(\omega) \\
&\quad + \frac{Q_1^2}{2C_{T,\text{eff}}} - ME_T \cos\left(2\pi\frac{(\phi_1 - \phi_J)}{M\phi_0}\right) + \frac{Q_J^2}{2C_J} - E_J \cos\left(2\pi\frac{\phi_J}{\phi_0}\right)
\end{aligned} \tag{22}$$

Here, $C_{T,\text{eff}} = C_cC_T/(C_c+C_T) = C_T/P$, where $C_c = 1/R_pK = \pi/RK$ is the equivalent of an additional capacitor in series with $C_T$, the magnitude of which depends on the bandwidth of the experimental setup, $K$ and the environmental impedance, R.

Now, we can quantize the Hamiltonian using the following relations by writing the coor-

---
[1] This leads to $K$ being limited to the order of the superconducting gap associated with the device metal

dinates in terms of ladder operators

$$q(\omega) = \sqrt{\frac{\hbar}{2}}(\hat{b}_\omega + \hat{b}^\dagger_\omega), \qquad p(\omega) = \sqrt{\frac{\hbar}{2}}\frac{(\hat{b}_\omega - \hat{b}^\dagger_\omega)}{i}, \tag{23}$$

$$\phi_1 = \sqrt{\frac{\hbar Z_T}{2}}(\hat{A}_T + \hat{A}^\dagger_T), \qquad Q_1 = \sqrt{\frac{\hbar}{2Z_T}}\frac{(\hat{A}_T - \hat{A}^\dagger_T)}{i}, \tag{24}$$

$$\phi_J = \sqrt{\frac{\hbar Z_J}{2}}(\hat{A}_J + \hat{A}^\dagger_J), \qquad Q_J = \sqrt{\frac{\hbar}{2Z_J}}\frac{(\hat{A}_J - \hat{A}^\dagger_J)}{i}. \tag{25}$$

In the above relations $Z_T = \sqrt{L_T/C_{T,\text{eff}}}$, $\Omega_{T,\text{eff}} = 1/\sqrt{L_T C_{T,\text{eff}}}$; $Z_J = \sqrt{L_J/C_J}$, $\Omega_J = 1/\sqrt{L_J C_J}$ are the impedance and frequencies of the linearised transformer and JPA oscillators respectively, where $L_T = \phi_0^2/4\pi^2 E_T$ and $L_J = \phi_0^2/4\pi^2 E_J$

$$\begin{aligned}\hat{H} &= \hat{H}_{Env} + \hat{H}_{Int} + \hat{H}_J + \hat{H}_T \\ &= \hat{H}_{Env} + \hat{H}_{Int} + \hat{H}_{\text{sys}}\end{aligned} \tag{26}$$

The second term in the above Hamiltonian is the coupling of the transformer to the environment. The coupling is proportional to $\sqrt{\omega}$. In the subsequent discussion, we have approximated that the coupling is equal to the decay rate of the device by removing the explicit dependence $\omega$ by replacing $\omega$ with $\Omega_{T,\text{eff}}$.

$$\begin{aligned}\hat{H} = {}& \hbar \int_0^\infty \omega \hat{b}^\dagger(\omega)\hat{b}(\omega)\,d\omega + \hbar \int_0^\infty \sqrt{\frac{\kappa}{2\pi}}(\hat{A}_T - \hat{A}^\dagger_T)(\hat{b}(\omega) - \hat{b}^\dagger(\omega))\,d\omega \\ & - \frac{\hbar\Omega_J}{4}(\hat{A}_J - \hat{A}^\dagger_J)^2 - E_J \cos\Big(c_2(\hat{A}_J + \hat{A}^\dagger_J)\Big) \\ & - \frac{\hbar\Omega_{T,\text{eff}}}{4}(\hat{A}_T - \hat{A}^\dagger_T)^2 - M^2 E_T \cos\left(c_1 \frac{(\hat{A}_T + \hat{A}^\dagger_T)}{M} - c_2 \frac{(\hat{A}_J + \hat{A}^\dagger_J)}{M}\right),\end{aligned} \tag{27}$$

where $\kappa = \frac{R\Omega_{T,\text{eff}}}{Z_T}$, is the linewidth of the transformer, and $c_1 = \frac{2\pi}{\phi_0}\sqrt{\frac{\hbar Z_T}{2}}$, $c_2 = \frac{2\pi}{\phi_0}\sqrt{\frac{\hbar Z_J}{2}}$ are introduced as dimensionless zero-point phase fluctuations across the transformer and the JPA.

## 3 Equations of Motion for the IEJPA

Armed with the full Hamiltonian of our system, now we try to write the full equations of motion for the IEJPA while trying to make the minimum number of assumptions and approximations in order to try and explain our obtained data.

Starting with the Hamiltonian 27, we notice that the linear coupling to the environment

lends the system quite well to an analysis using input-output theory. Under the input-output formalism, the quantum Langevin equation for intracavity fields for the transformer and the JPA follow [4]

$$\frac{d\hat{A}_T}{dt} = -\frac{i}{\hbar}[\hat{A}_T, \hat{H}_{\text{sys}}] - \frac{\kappa}{2}\hat{A}_T + \sqrt{\kappa}\hat{A}_{\text{in}}, \quad (28)$$

$$\frac{d\hat{A}_J}{dt} = -\frac{i}{\hbar}[\hat{A}_J, \hat{H}_{\text{sys}}]. \quad (29)$$

For the amplifier mode of operation of the IEJPA, one injects a strong classical pump from the environment into the system displacing the intracavity fields. We then look at the modified Hamiltonian or equations of motion to study the dynamics of a weak, perturbative, quantum signal in the presence of these coherent classical displacements. To capture these dynamics, we decompose the intracavity fields $\hat{A}_T, \hat{A}_J$ into $\hat{A}_T(t) = (\alpha_T + \hat{a}_T(t))e^{-i\omega_p t}, \hat{A}_J(t) = (\alpha_J + \hat{a}_J(t))e^{-i\omega_p t}$ where $\alpha_T, \alpha_J$ are classical pumps, and $\hat{a}_T, \hat{a}_J$ are the intracavity operators encoding the perturbations about these classical displacements inside the transformer and JPA.

First, as the full Hamiltonian is abundantly nonlinear, we list out the two commutation relations of interest in terms of $\hat{A}_T, \hat{A}_J$. We write

$$[\hat{A}_T, \hat{H}_{\text{sys}}] = \frac{\hbar\Omega_{T,\text{eff}}}{2}(\hat{A}_T - \hat{A}_T^\dagger) + ME_T c_1 \sin\left(c_1\frac{(\hat{A}_T + \hat{A}_T^\dagger)}{M} - c_2\frac{(\hat{A}_J + \hat{A}_J^\dagger)}{M}\right), \quad (30)$$

$$[\hat{A}_J, \hat{H}_{\text{sys}}] = \frac{\hbar\Omega_J}{2}(\hat{A}_J - \hat{A}_J^\dagger) + E_J c_2 \sin\left(c_2(\hat{A}_J + \hat{A}_J^\dagger)\right) - ME_T c_2 \sin\left(c_1\frac{(\hat{A}_T + \hat{A}_T^\dagger)}{M} - c_2\frac{(\hat{A}_J + \hat{A}_J^\dagger)}{M}\right). \quad (31)$$

In these expressions, it is the most tedious to handle the sine-nonlinearities. To resolve them, we first split the operators above into the classical and quantum parts and then assuming that the quantum fluctuations are very small ($\langle\hat{a}^\dagger\hat{a}\rangle \cdot \max(c_1, c_2)/M \ll 1$), expand the sines into a purely classical sine part of the pump and a term that couples the pump in

7

a cosine term to a linear term in $\hat{a}$. To illustrate, we write

$$\sin\left(c_1\frac{(\hat{A}_T+\hat{A}_T^\dagger)}{M} - c_2\frac{(\hat{A}_J+\hat{A}_J^\dagger)}{M}\right)$$

$$= \sin\left(\frac{(c_1\alpha_T - c_2\alpha_J)e^{-i\omega_p t} + (c_1\alpha_T^* - c_2\alpha_J^*)e^{i\omega_p t}}{M} + \frac{(c_1\hat{a}_T - c_2\hat{a}_J)e^{-i\omega_p t} + (c_1\hat{a}_T^\dagger - c_2\hat{a}_J^\dagger)e^{i\omega_p t}}{M}\right)$$

$$= \sin\left(\frac{2|c_1\alpha_T - c_2\alpha_J|}{M}\cos(\omega_p t - \phi_{\text{eff}}) + \frac{(c_1\hat{a}_T - c_2\hat{a}_J)e^{-i\omega_p t} + (c_1\hat{a}_T^\dagger - c_2\hat{a}_J^\dagger)e^{i\omega_p t}}{M}\right) \quad (32)$$

$$= \sin\left(\frac{2|c_1\alpha_T - c_2\alpha_J|}{M}\cos(\omega_p t - \phi_{\text{eff}})\right)$$
$$+ \frac{(c_1\hat{a}_T - c_2\hat{a}_J)e^{-i\omega_p t} + (c_1\hat{a}_T^\dagger - c_2\hat{a}_J^\dagger)e^{i\omega_p t}}{M}\cos\left(\frac{2|c_1\alpha_T - c_2\alpha_J|}{M}\cos(\omega_p t - \phi_{\text{eff}})\right), \quad (33)$$

where $\phi_{\text{eff}}$ is defined as $(c_1\alpha_T - c_2\alpha_J) = |(c_1\alpha_T - c_2\alpha_J)|e^{i\phi_{\text{eff}}}$. Similarly, to express the pump in the JPA's nonlinear potential, we define $\phi_{\text{jpa}}$ as $c_2\alpha_J = |c_2\alpha_J|e^{i\phi_{\text{jpa}}}$

Under this expansion, one can now write the equations of motion of the classical pump first assuming $\hat{a}_J, \hat{a}_T \to 0$, and then substituting the obtained steady-state occupancies of the pump to solve for the quantum behaviour of the system under a weak signal.

## 3.1 Classical Equations of Motion for the Pumps

The equations of motion for the two pump displacements are given by

$$\dot{\alpha}_T e^{-i\omega_p t} = i\omega_p \alpha_T e^{-i\omega_p t} - \frac{\kappa}{2}\alpha_T e^{-i\omega_p t} + \sqrt{\kappa}\alpha_{\text{in}}e^{-i\omega_p t}$$
$$- \frac{i}{\hbar}\left(\frac{\hbar\Omega_{T,\text{eff}}}{2}(\alpha_T e^{-i\omega_p t} - \alpha_T^* e^{i\omega_p t}) + ME_T c_1 \sin\left(\frac{2|(c_1\alpha_T - c_2\alpha_J)|}{M}\cos(\omega_p t - \phi_{\text{eff}})\right)\right), \quad (34)$$

$$\dot{\alpha}_J e^{-i\omega_p t} = i\omega_p \alpha_J e^{-i\omega_p t} - \frac{i}{\hbar}E_J c_2 \sin(2|c_2\alpha_J|\cos(\omega_p t - \phi_{\text{jpa}}))$$
$$- \frac{i}{\hbar}\left(\frac{\hbar\Omega_J}{2}(\alpha_T e^{-i\omega_p t} - \alpha_T^* e^{i\omega_p t}) - ME_T c_2 \sin\left(\frac{2|c_1\alpha_T - c_2\alpha_J|}{M}\cos(\omega_p t - \phi_{\text{eff}})\right)\right). \quad (35)$$

Finally, we assume that $\omega_p$ is much larger than all the detunings and rates of dynamics in our system. Under this assumption, we can make the rotating wave approximation, and also capture the exact effects of the pump up to the validity of the RWA by making use of the Jacobi-Anger transformation[5]. Using the Jacobi-Anger expansion, we can get the exact dynamics at a frequency $\omega_p$. Performing the expansion, applying the RWA, and simplifying,

we obtain the classical equations of motion of the pump fields

$$\dot{\alpha}_T = \left(i\omega_p - i\frac{\Omega_{T,\text{eff}}}{2} - \frac{\kappa}{2}\right)\alpha_T + \sqrt{\kappa}\alpha_{\text{in}} - i\frac{ME_T}{\hbar}c_1 J_1\left(\frac{2|c_1\alpha_T - c_2\alpha_J|}{M}\right)e^{i\phi_{\text{eff}}}, \tag{36}$$

$$\dot{\alpha}_J = \left(i\omega_p - i\frac{\Omega_j}{2}\right)\alpha_J + i\frac{ME_T}{\hbar}c_2 J_1\left(\frac{2|c_1\alpha_T - c_2\alpha_J|}{M}\right)e^{i\phi_{\text{eff}}} - i\frac{E_J}{\hbar}c_2 J_1\left(2|c_2\alpha_J|\right)e^{i\phi_{\text{jpa}}}. \tag{37}$$

An extremely important feature of these equations of motion is the existence of multiple stable steady-state solutions, i.e. solutions such that $\dot{\alpha}_J = \dot{\alpha}_T = 0$ and the stable points are attractors. This makes solving for the state the classical pump reaches hard. To solve this problem, we solve the differential equations starting from $\alpha_J(t=0) = \alpha_T(t=0) = 0$, and then evolving the system for a long time followed by performing a local root-finding for $\dot{\alpha}_J = \dot{\alpha}_T = 0$.

## 4  Signal Equations For Transformer and JPA

Now that we have obtained the steady state intracavity pump fields $(\alpha_T, \alpha_J)$, we can compute the perturbations due to the additional terms in the equations of motion Eqn. 28 and Eqn. 29. Under the RWA, applying the Jacobi-Anger expansion, we get the quantum equations of motion

$$\begin{aligned}\frac{d\hat{a}_T(t)}{dt} &= (i\omega_p - i\frac{\Omega_{T,\text{eff}}}{2} - \frac{\kappa}{2})\hat{a}_T(t) + \sqrt{\kappa}\hat{a}_{\text{in}}(t) \\ &\quad - i\frac{E_T}{\hbar}c_1 J_0(A_{\text{eff}})(c_1\hat{a}_T(t) - c_2\hat{a}_J(t)) \\ &\quad + i\frac{E_T}{\hbar}c_1 J_2(A_{\text{eff}})\left(c_1\hat{a}_T^\dagger(t) - c_2\hat{a}_J^\dagger(t)\right)e^{i2\phi_{\text{eff}}},\end{aligned} \tag{38}$$

$$\begin{aligned}\frac{d\hat{a}_J(t)}{dt} &= i\left(\omega_p - \frac{\Omega_J}{2}\right)\hat{a}_J(t) + i\frac{E_T}{\hbar}c_2 J_0\left(A_{\text{eff}}\right)\left(c_1\hat{a}_T(t) - c_2\hat{a}_J(t)\right) \\ &\quad - i\frac{E_T}{\hbar}c_2 J_2\left(A_{\text{eff}}\right)\left(c_1\hat{a}_T^\dagger(t) - c_2\hat{a}_J^\dagger(t)\right)e^{i2\phi_{\text{eff}}} \\ &\quad - i\frac{E_J}{\hbar}c_2^2 J_0\left(A_{\text{jpa}}\right)\hat{a}_J(t) + i\frac{E_J}{\hbar}c_2^2 J_2\left(A_{\text{jpa}}\right)\hat{a}_J^\dagger(t)e^{i2\phi_{\text{jpa}}}.\end{aligned} \tag{39}$$

where, $A_{\text{eff}} = 2|c_1\alpha_T - c_2\alpha_J|/M$, $A_{\text{jpa}} = 2|c_2\alpha_J|$.

From these quantum Langevin equations, the gain-spectrum of the IEJPA can be obtained by expressing them in the frequency domain, while noting that the perturbation operators are represented in a frame rotating with the pump. So, for a signal with the lab-frame frequency $\omega_s$, the Fourier transformed equations would give a solution at $\Delta = \omega_s - \omega_p$.

With this, we express,

$$-i\Delta\hat{a}_T(\Delta) = \left(i\omega_p - i\frac{\Omega_{T,\text{eff}}}{2} - \frac{\kappa}{2}\right)\hat{a}_T(\Delta) + \sqrt{\kappa}\hat{a}_{\text{in}}(\Delta)$$
$$- i\frac{E_T}{\hbar}c_1 J_0(A_{\text{eff}})(c_1\hat{a}_T(\Delta) - c_2\hat{a}_J(\Delta))$$
$$+ i\frac{E_T}{\hbar}c_1 J_2(A_{\text{eff}})\left(c_1\hat{a}_T^\dagger(-\Delta) - c_2\hat{a}_J^\dagger(-\Delta)\right)e^{i2\phi_{\text{eff}}}, \quad (40)$$

$$-i\Delta\hat{a}_J(\Delta) = i\left(\omega_p - \frac{\Omega_J}{2}\right)\hat{a}_J(\Delta) + i\frac{E_T}{\hbar}c_2 J_0(A_{\text{eff}})(c_1\hat{a}_T(\Delta) - c_2\hat{a}_J(\Delta))$$
$$- i\frac{E_T}{\hbar}c_2 J_2(A_{\text{eff}})\left(c_1\hat{a}_T^\dagger(-\Delta) - c_2\hat{a}_J^\dagger(-\Delta)\right)e^{i2\phi_{\text{eff}}}$$
$$- i\frac{E_J}{\hbar}c_2^2 J_0(A_{\text{jpa}})\hat{a}_J(\Delta) + i\frac{E_J}{\hbar}c_2^2 J_2(A_{\text{jpa}})\hat{a}_J^\dagger(-\Delta)e^{i2\phi_{\text{jpa}}}. \quad (41)$$

The equations above, along with their hermitian conjugates, can be used to solve the relations between the input and output fields using the relation [2]

$$\sqrt{\kappa}\hat{a}_T(\Delta) = \hat{a}_{\text{in}}(\Delta) + \hat{a}_{\text{out}}(\Delta). \quad (42)$$

Finally, for linear systems, the only solution to the relation between the input and output fields would be of the form

$$\hat{a}_{\text{out}} = \sqrt{G(\Delta)}\hat{a}_{\text{in}}(\Delta) + \sqrt{G(\Delta)-1}\hat{a}_{\text{in}}^\dagger(-\Delta), \quad (43)$$

constrained by the commutation relations between bosonic operators describing a travelling wave.


\clearpage

\end{document}

%% file: main_v3.bbl
%